# A distribution-dependent analysis of open-field test movies

Tomokazu Konishi, and Haruna Ohrui

Graduate School of Bioresource Sciences, Akita Prefectural University

**ABSTRACT**

Although the open-field test has been widely used, its reliability and compatibility are frequently questioned. Although many indicating parameters were introduced for this test, they did not take data distributions into consideration. This oversight may have caused the problems mentioned above. Here, an exploratory approach for the analysis of video records of tests of elderly mice was taken that described the distributions using the least number of parameters. First, the locomotor activity of the animals was separated into two clusters: dash and search. The accelerations found in each of the clusters were distributed normally. The speed and the duration of the clusters exhibited an exponential distribution. Although the exponential model includes a single parameter, an additional parameter that indicated instability of the behaviour was required in many cases for fitting to the data. As this instability parameter exhibited an inverse correlation with speed, the function of the brain that maintained stability would be required for a better performance. According to the distributions, the travel distance, which has been regarded as an important indicator, was not a robust estimator of the animals' condition.

## 1. Introduction

The open-field test has been used to investigate the locomotor activity of animals. The animals' behaviour is summarized using sets of indicating parameters [1-7]. However, questions repeatedly arise regarding the reliability and compatibility of the open-field test given the variety of test apparatuses and procedures [1, 5, 6]. In recent years, video-recording systems have become popular [3, 4, 8] because they can accurately capture animal behaviour; however, the movies comprise a large volume of complex information. It may be possible to summarize such information systematically using exploratory data analysis to identify the patterns of data distribution, which has been disregarded in the past [9].

The distribution model can describe the characteristics of the data using the least number of parameters. This approach has several advantages; e.g., the parameters can describe the characteristics of the data. The distribution may provide clues about the data's hidden background mechanism, which determines the nature of the data. Because the distribution pattern may be a common character among experiments, it may provide a standard unit that can be shared beyond experiments, as is the case for $z$-scores. Moreover, the character of the distribution will indicate which parameter is sufficiently robust to overcome the experimental noise. Thus, the distribution model would improve the reliability and compatibility of the analysis.

## 2. Materials and methods

Four female C57BL/6NJcl mice were obtained from CLEA Japan, Inc. (Tokyo, Japan). They were housed together in a room with controlled temperature (22°C ± 2°C) and a 12 h light/dark cycle. The open-field tests were performed at 60–110 weeks of age in a white polypropylene container (30.9 × 43.9 cm; height, 30 cm) in the daytime without altering the daily light conditions. The activity was recorded as a movie using a digital camera (IXY 430F; Canon, Tokyo, Japan). The present study was approved by the Review Board of the Animal Ethics Committee of the Akita Prefectural University (H19-04) and followed institutional guidelines for the care and use of laboratory animals.

The condition of the test animals at a given moment was estimated as follows. The recorded movie was read as a stack of images using the ImageJ program [10]. Each of the images was binarized to determine the shape of the animal; subsequently, the area and position of the centre of mass were determined. The table of areas and positions was transferred and analysed using R software [11]. The detailed protocols and scripts used for the movie analysis are included in the Mendeley Data repository.

## 3. Results

### 3.1. Movement of the centre of mass in the body shape

The animals engaged in activity constantly without rest during the 30 min of each test, which was in contrast with their ordinary diurnal behaviour; i.e., sleeping. As is well known, the animals rarely entered the central region of the field; rather, they moved in the peripheral space (Fig. 1A). The speed of their movement was not constant; e.g., when an animal went around the field for 10 s (Fig. 1B and Mendeley Data movie 1), the recorded speed exhibited five sharp peaks (Fig. 1C, solid line). The accelerations and decelerations also exhibited peaks (dotted line), which appeared at half of the peak speed. Between movements, the animals searched around; i.e., they moved slowly and engaged in undirected sniffing, raised their forelimbs when rearing, or investigated the walls (to escape from the container).

*3.2. Shape of the area of animal movement*

Viewed from above, the shape of the areas of the animal movements provided information about their habitual movements. The area records showed a tendency to be smaller toward the end of the test (Fig. S1A), which was an artefact caused by drift of the exposure of the camera. Thus, a corrected data set was used for further analysis (Fig. S1B).

When the animals ran fast, their stretched bodies showed larger areas. During rearing, the areas became smaller. Hence, the area and speed of the animals exhibited a weak correlation (Fig. 1D). The plot exhibited two clusters; i.e., "faster and larger" and "slower and smaller" clusters. Although those clusters should show some overlap, for simplicity, the clusters were separated using a line. The slope and intercept of the separating line were temporarily placed manually by the analysist and were then adjusted during the fitting process of the speed distributions, as described below.

For descriptive purposes, we called the two clusters dash and search, respectively. Both clusters appeared sequentially as a series array in the tracking records (Fig. 1B, 1C).

*3.3. Noise*

The original record of the centre mass area contains noise. Any movement changes the recorded shape of the animal, which alters the estimations of its centre mass, even if the animal stays at the same location. Such alteration produces noise, which can be reduced by taking moving averages of the animal positions. The speed and acceleration data shown in Fig. 1A and C underwent such noise reduction; otherwise, it would have been difficult to analyse the raw record (Fig. S1C). The effect of noise reduction was larger at slower speeds (Fig. S1D).

*3.4. Data distribution*

If the data distribution obeys certain rules, they would show a pattern that could be described using a limited number of parameters, which represent the characteristics of data. Here, such patterns and parameters were introduced regarding the locomotor activity of the animals (Fig. 2). As the noise reduction would alter the distribution, the patterns were extracted directly from the raw data.

*3.5. Acceleration*

The accelerations were approximated using three normal distributions. The normal distribution has two parameters: location ($\mu$) and scale ($\sigma$). In this case, the three normal distributions had their own mixing ratios of the whole. All distributions had $\mu = 0$ (Fig. 2A); therefore, the three accelerations could be practically described using six parameters.

The largest $\sigma$ was found for accelerations of the dash cluster (Fig. S2A). The quantile–quantile (Q–Q) plot compares the sorted data against the normal distribution; i.e., if the data are normally distributed, the plot will produce a straight line. The minor distribution with the smallest $\sigma$ might reflect the noise of the raw record. An intermediate $\sigma$ was found by separating the noise fraction from the search cluster. The normality of the distribution indicated that the accelerations and decelerations had the same magnitude.

*3.6. Speed*

The logarithms of the speed were approximated using a mix of four normal distributions: two major and two minor ones (Fig. S2B). However, a better fitting was achieved using an exponential distribution (Fig. 2B). The fine-tuning process was performed by determining the line that separates the dash and search clusters (Fig. 1D). The other parameters were automatically found in the linear relationships with the speed records of the dash or search clusters in a log scale (Fig. S2C, S2D).

The exponential distribution has only one parameter, $\lambda$, which determines the rate of occurrence (Mendeley Data, see Appendix); its probability density function is monotonically decreasing, and the density was concentrated near zero (Fig. S2E). Such heavy skewing could reduce the accuracy of parameter estimations; however, the skewing is eased by taking the logarithms of the speed (Fig. 2B). Here, the data distribution of log(Speed) was approximated by a linear transformation of the logarithms of two sets of exponential distributions ($\lambda = 1$) (Fig. S2C, S2D). Hence, the number of parameters used was three for both the dash and search clusters: mixing rates, slopes *a*, and intercepts *b* (Mendeley Data, see Appendix). For estimations of the parameters *a* and *b*, random sampling gave errors that could be approximated by normal distributions (Fig. S2F).

The exponential distribution represents intervals of events that occur randomly with a constant rate, $\lambda$. The new parameter *a* changes the shape of the distribution (Appendix). When *a*<1, the distribution is concentrated around its mode. When the animal did not keep the rate $\lambda$, the distribution exhibited a wider scale and *a* became >1 (Fig. 2C). In short, parameter *a* is indicative of the instability of the behaviour.

*3.7. Area*

The body areas found in the dash and search clusters did not show a rigid pattern, suggesting that the estimation of areas may not be accurate and may contain a larger level of noise than that of speed. However, the bell shapes could be approximated by normal distributions (Fig. S2G). The video showed that a value 0.6 times $\sigma$ lower than $\mu$ seemed to be indicative of rearing (Mendeley Data, movie 2 shows examples of the observed rearing).

*3.8. Distribution of the duration of the forms of behaviour*

After separating the forms of behaviours into dash and search (or rearing and grounding: keeping the four paws on the ground), their duration became clear (Figs 2D, S2H). Their logarithms were approximated by logarithms of the exponential distribution, as was the case for speed (Fig. S2I–L). The distribution of the model exhibited a slight skew toward a negative value and gave a better fit than those that were obtained using a lognormal distribution without skewing (Figs 2D, S2H).

*3.9. Relationships among the parameters*

Some parameters showed correlations with each other. As expected, the modes of acceleration and speed of the dash set were correlated (Fig. 3A). Animals with better stability were faster (Fig. 3B). The instability parameters *a* showed correlation among the various forms of behaviour (Fig. S3A–C).

Because the speed and frequency of the dash cluster showed a weak inverse correlation (Fig. S3D), the faster animals did not always travel farther (Fig. 3C). One of the tested mice showed a decrease in dash speed; this back and forth between the two phases may be an indication of aging (Fig. 3D). This was accompanied by the increase of the *a* parameter of speed (Fig. S3E). This phenomenon was not apparent for the travel distance (Fig. S3F).

**4. Discussion**

A movie contains a large volume of information that guarantees accurate and objective analyses. Finding the specific distributions and their parameters is beneficial for summarizing the data and extracting the characteristics of the test animals.

One of the benefits of knowing the data distribution becomes obvious during the investigation of rearing. The data obeyed the exponent of the exponential distribution. Data handling without this knowledge is quite difficult, for the following reasons: first, it is required for identifying the parameters. Most of the duration of behaviours is concentrated within very short periods. This hampers observations; thus, compensations for counting omissions or false positives are indispensable, even just for assessing frequency. Here, the compensation was performed by comparing the data quantiles with the exponential distribution (Fig. S2K, L). Second, the identified parameters tell the state of animals. The heavily skewed distribution renders the arithmetic mean useless because it is not robust and does not show the mode. Therefore, measuring the duration repeatedly using a manual approach and taking the mean does not generate useful information. In contrast, the parameters *a* and *b* can describe the behaviour of the animal accurately and can be treated parametrically (Fig. S2F). Those parameters will inform on the frequency, length, and constancy of the recorded behaviours. Because of the distribution model, those parameters can be extracted from movie records automatically. In contrast, the naked eye cannot follow the quick movement of mice.

The modes of the steady distributions, such as those observed for speed, should be robust indicators of the animals' locomotor activities (Fig. 3D). Conversely, it has become clear that the distance of movement is not a robust indicator, as the distance is the sum total of the product of speed and duration, the distributions of which are heavily skewed (Fig. 2B, D). In a mathematical sense, it is certain that the product is unstable and can be easily altered by any tiny effects that occur by chance. In turn, the distance did not indicate the vitality of the animal in a linear way. In fact, the fastest animal did not always travel farther (Fig. 3C) and the effect of aging was not observed in the case of distance (Fig. S3F).

The contraction of muscles may explain the distribution pattern of accelerations. According to Newton's equation of motion, acceleration is directly proportional to the force applied. The strength of the force given by the muscle of the test animal is proportional to the number of muscle cells that are activated at that moment. At a specific moment, the selection of activated muscle cells, which were stimulated by the excited nerves, would exhibit a certain randomness. If this is true, the number of activated cells would be normally distributed, as observed here.

It should be noted that accelerations and decelerations showed the same magnitudes (Figs 2A and S2A). This finding shows that, although they depend on different sets of muscles, the same amount of force was consumed. The deceleration was not the follow-through of a running event, but rather, it was active speed braking. This clearly shows that activity is controlled to satisfy the animals' purposes.

The mode of dash speed obeyed the exponentiation of the exponential distribution (Figs 2B and S2C). This reflected the distribution of the duration of the dash cluster (Fig. 2D). As the acceleration was normally distributed (Figs 2A and S2A), the speed should also be normally distributed if it was just a sum of randomly selected accelerations. However, the accelerations were coherently coordinated with the speed (Fig. 1C). In many cases, the instability parameter $a$ was nearly 1 (Figs 2D and S2I–L) and exhibited an ordinal exponential distribution, which may indicate intervals of randomly occurring events. However, in some cases, compensation by using the instability parameter $a$ was required to approximate the data distribution. The parameter changes the distribution to either concentrating on the mode or instability of $\lambda$.

A smaller $a$ value may indicate a better condition of the brain. Although the smaller $a$ observed for duration in the dash cluster does not mathematically guarantee a faster speed (e.g., a longer dash duration could produce a faster speed), animals with more stable behaviour were faster in reality (Fig. 3B). Moreover, faster animals showed larger accelerations, which require the coincidental activation of a larger number of muscle cells (Fig. 3A). These observations imply that a stricter control by the brain is required for achieving a faster speed. In the case of a smaller $a$, the distribution would be concentrated around their modes. Therefore, the peak speed and duration values in a test become constant. As the faster speed guarantees a shorter arrival time, the duration and frequency of dash were decreased (Fig. S3A, B, D), which favours safer movement. Hence, the constant pattern of behaviour does not imply a stereotypical behaviour; rather, repeating a suitable movement within a box would generate such a steady pattern. In contrast, the loss of this ability increased the instability parameter *a*; eventually, the occurrence rate of behaviours, $\lambda$, would become unstable (Fig. 2C). As the *a* values found in different forms were correlated (Fig. S3A–C), the same or related parts of the brain seem to maintain the stability of behaviours (Fig. 2C).

Although separating the dash from search cluster (Fig. 1D) worked well for estimating the parameters, it may have produced a sort of error. As the two clusters are closely located, they must have some overlap; the same problem also exists in the separation of rearing and grounding (Fig. S2G). Therefore, classifying a moment into the two forms has a basic difficulty. The error will cut several continuous actions into smaller pieces, or overlook shorter fragments. This would be the cause of the departure at the lower part of Q–Q plots (Fig. S2 C, D, I–L). Hence, the slope and intercept parameters should be extracted from the upper part of the plots.

The exponential distribution and the lognormal distribution are similar; thus, either of them can be used to approximate the distributions of speed or duration. However, the former reproduced the mode with better fidelity (Figs 2B and S2B, 2D, S2H); this characteristic will improve the accuracy of parameter estimations that use fitting processes. Moreover, the distribution model may become a hint of the hidden structure of the data [9, 12], which generates it, such as that shown for the mechanism of transcriptome production [13]. Here, it was difficult to explain the lognormal distributions, but a tentative explanation of the exponential distribution is available.

As the sampling noise was practically distributed normally (Fig. S2F), the parameters have potential compatibility with parametric models, such as ANOVA; this is beneficial for statistical tests. However, it should be noted that we must pay attention to the individual differences among animals; these would be especially apparent for older animals. In such cases, it is not adequate to expect a normal distribution of the experimental noise; rather, an experimental design that considers this issue will be required, such as pairwise comparisons.

The distributions and the parameters can be easily extracted from video records. A parametric analysis performed by checking the distributions will improve the accuracy of the analysis. According to this point of view, the traditional methods of estimation of moving distance, such as those that use grid beam breaking, may have a high level of noise, and the distance itself is not a robust indicator of the animal's state. Such inappropriate choices of indicating parameters may have been at the origin of the concerns raised regarding the open-field tests [1, 5, 6].

# Figures

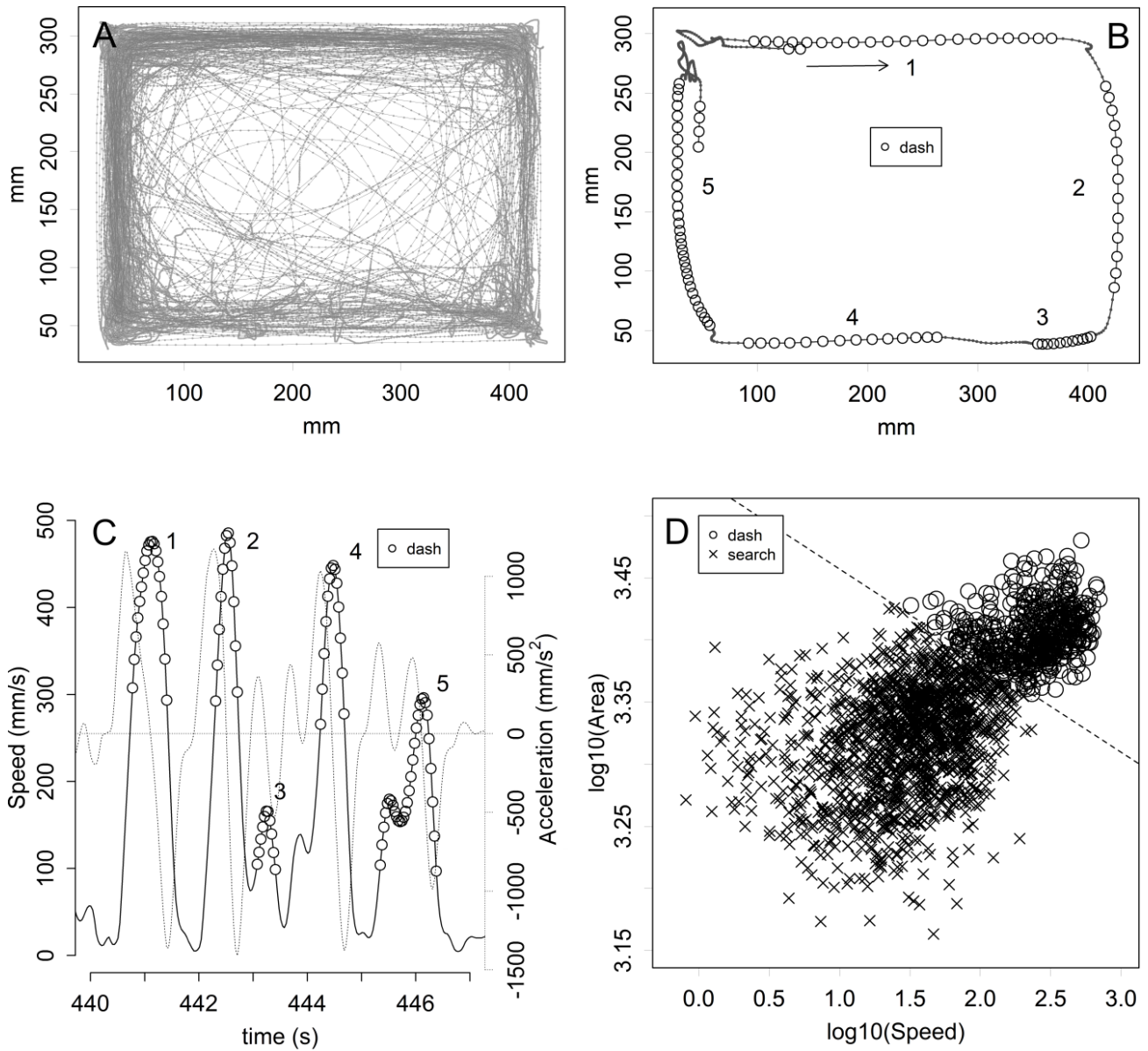

**Fig. 1. A**. Tracking record of a mouse during a 30-min test. The dots show the position of the centre of mass in each image; the time interval of the dots was 1/30 s. **B**. Part of the tracking record, compassing the field for 10 s. The sequential circles indicate the dash time, which is separated by the searching times (dots). **C**. Changes in the speed and acceleration of the compassing. The speed showed five sharp peaks (solid line). Each peak was wedged with positive and negative peaks of acceleration (dotted line). **D**. Comparison of speed and area. A weak correlation was produced by the two clusters that are separated by a straight dotted line.

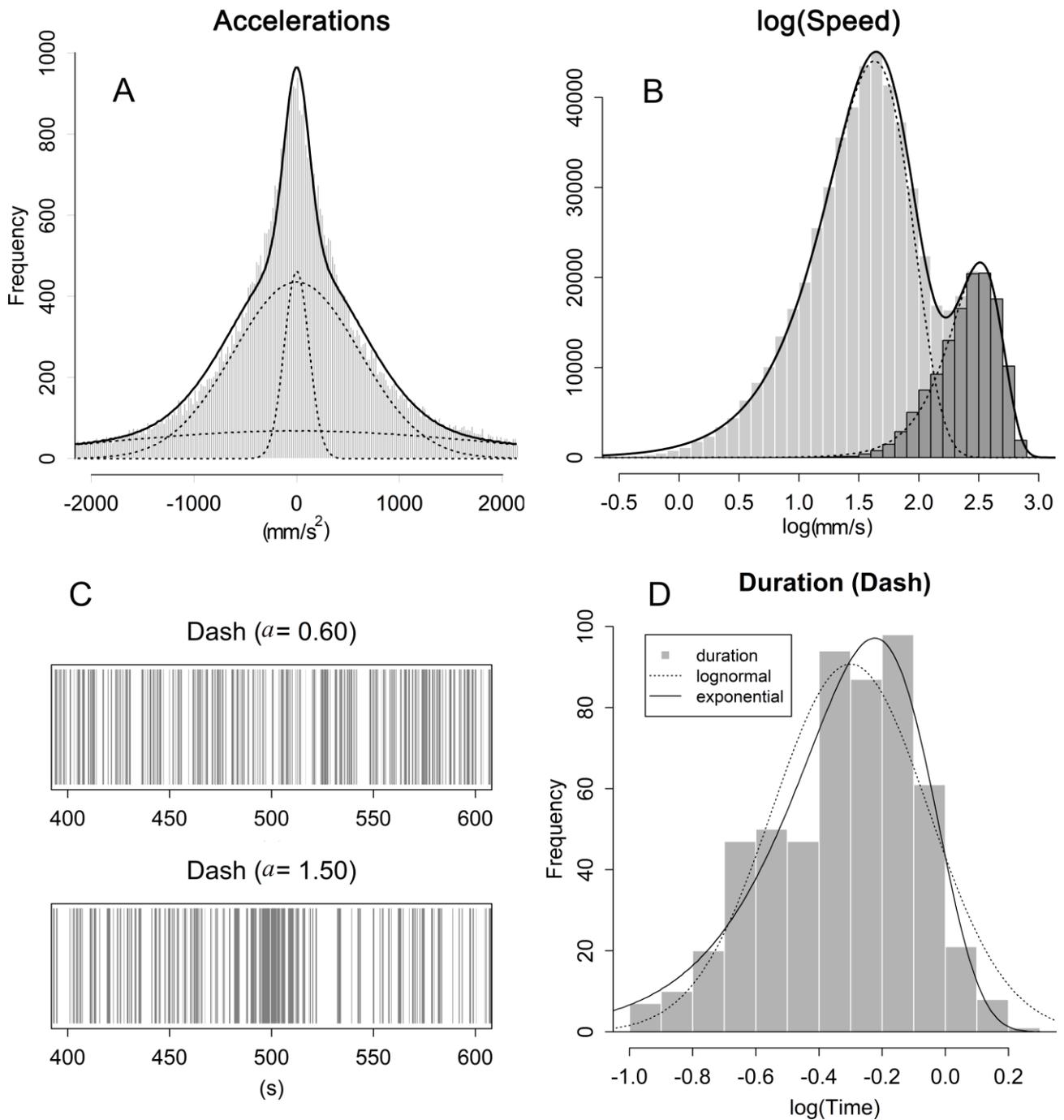

**Fig. 2**. **A**. Distribution of the accelerations. The sum of three normal distributions fit the distribution. The three distributions have zero mean and different $\sigma$. The largest $\sigma$ was observed for the dash cluster (Fig. S2A). **B**. Distribution of the logarithms of speed. Fitting results obtained using two exponential distributions are shown (dotted lines; cf, fitting using lognormal distributions, Fig. S2B). The darker area represents the dash cluster. **C**. Dash cluster times are indicated as grey vertical bars. When *a* is <1, the width of the bar is homogeneous (upper panel), and vice versa. The intervals show the duration of the search, as well as the differences in homogeneity, as the *a* values in different forms of behaviour were correlated (Fig. S3B). **D**. Duration (log-scale) of the dash clusters. Fittings using a lognormal distribution (dotted line) and exponential distribution (solid line, log-scale) are shown. See distribution of the search cluster in Fig. S2H.

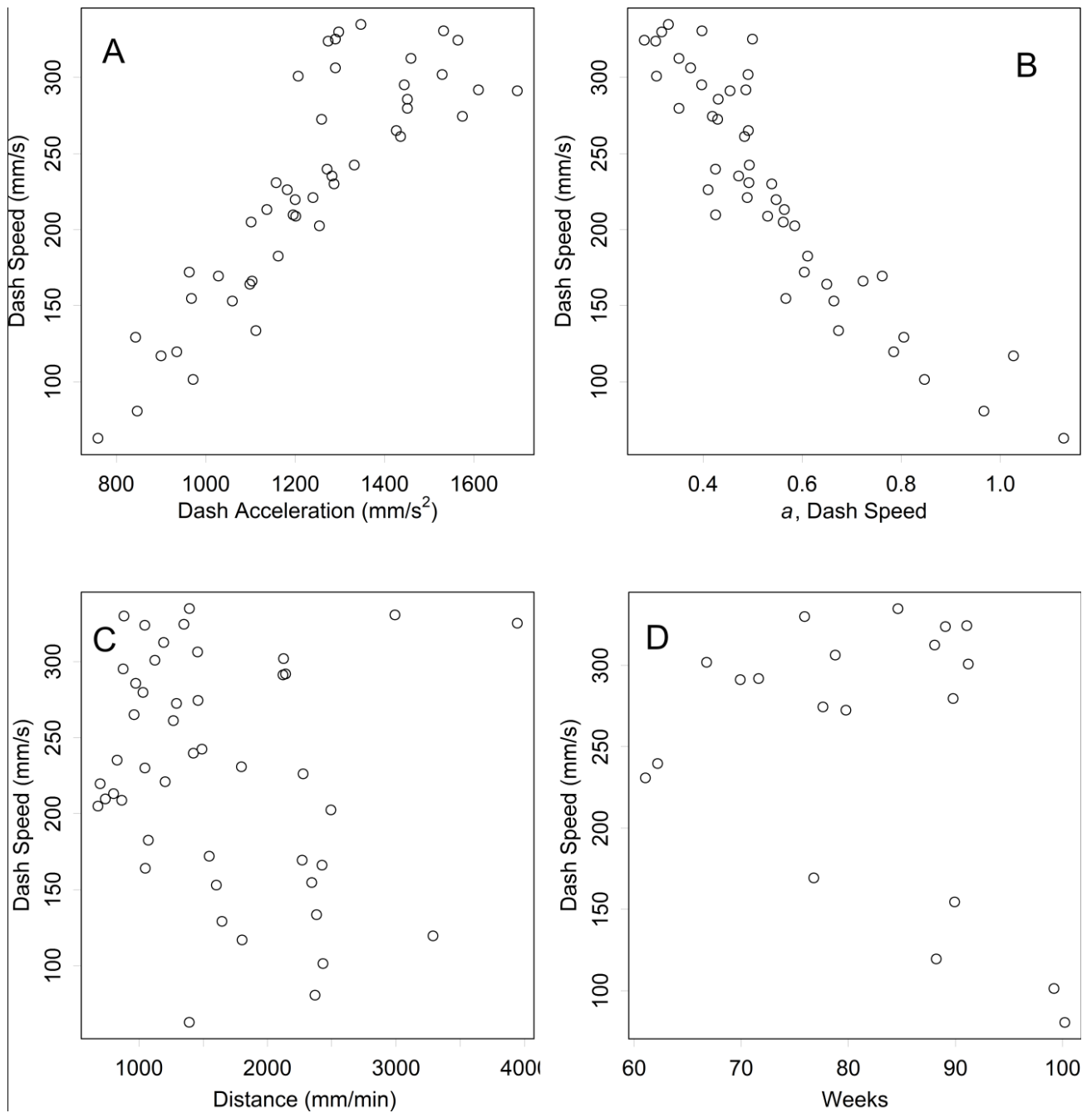

**Fig. 3**. **A**. Comparison of the mode of speeds and accelerations in the dash cluster found in each of the long periods of the tests. Each dot indicates the mode of a test. **B**. Comparison of the mode and $a$ of the dash speeds. **C**. Comparison of the dash speed mode and distances per min. **D**. Annual changes in the dash speed mode.

**Supplementary Data**

**Appendix**

*1. Exponentiation of the exponential distribution*

This simple model has only one rate parameter, $\lambda$, and the probability density function is:

$$f(x;\lambda) = \begin{cases} \lambda \exp(-\lambda) & x \geq 0 \\ 0 & x < 0 \end{cases} \quad \text{(Fig. S2E)},$$

where $x$ is the quantile. The quantiles and the cumulative probability $P$ have the following relationship:

$$x = \{-\ln(1 - P)\}/\lambda,$$

where the mean and the standard deviation are both $1/\lambda$. However, as the distribution is heavily skewed, these two parameters do not represent the characteristics of the distribution in a practical sense. Taking the logarithms of the quantiles, $\log(x)$, may ease the skewing and the distribution may become bell shaped (Fig. S4A). The rate parameter changes the location of the peak, $\log(1/\lambda)$, but not the width of the bell curves.

*2. Linear fitting to the logarithmic data*

Here, linear relationships were observed between the logarithms of the data and the exponential distribution ($\lambda = 1$), which implies that the slope was not always equal to 1. Applying the transformation leads to the exponentiation of quantiles, as follows:

$$a \times \log(x) + b = \log(x^a \times 10^b),$$

where $a$ is the slope and $b$ is the intercept. This changes the width of the log distribution (Fig. S4B) and the shape of the antilog distribution; when the exponent $a<1$, the density becomes bell shaped (Fig. S4C). This character was observed in the real data (Fig. S4D).

The altered probability density function is as follows. The quantiles of distribution, $x$, were determined as the exponentiation of the exponential distribution:

$$x = \left(\frac{-\ln(1-P)}{\lambda}\right)^a.$$

Solving this for $P$ gives the cumulative distribution function $F(x; a, \lambda) = P = 1 - \exp(-\lambda x^{1/a})$. This derives the density function:

$$f(x; a, \lambda) = \lambda/a \times x^{(\frac{1}{a}-1)} \exp(-\lambda x^{1/a}).$$

Hence, the distribution model has two parameters, $\lambda$ and $a$. Note that the expressions are slightly different from the well-studied exponentiated exponential distribution, which was derived from the exponentiation of $P$, but not $x$ [14]. Moreover, they are not derivatives of the gamma distribution [12].

Although the functions might not be familiar ones, in practice, the parameters can be determined robustly from the linear relationship in the log(Q–Q) plot of data against the exponential distribution ($\lambda = 1$). The expectation of the $a$ is the slope, and $\lambda = 10^{-b/a}$, where $b$ is the intercept. A value of $a = 1$ represents the exponent distribution, and $x$ will show the intervals of randomly occurring events. When $a<1$, the data $x$ will concentrate around the mode (Fig. S4B). When $a>1$, a stable $\lambda$ cannot be expected (Fig. 2D).

The peak of the probability density function represents the mode of the log data at $\log(x) = b = \log(1/\lambda^a)$ (Fig. 4B). As the log distribution is skewed, the peak is not equal to the mean of the log data. Moreover, the peak appears at $P = 1 - 1/e \approx 0.632$, which is not the median.

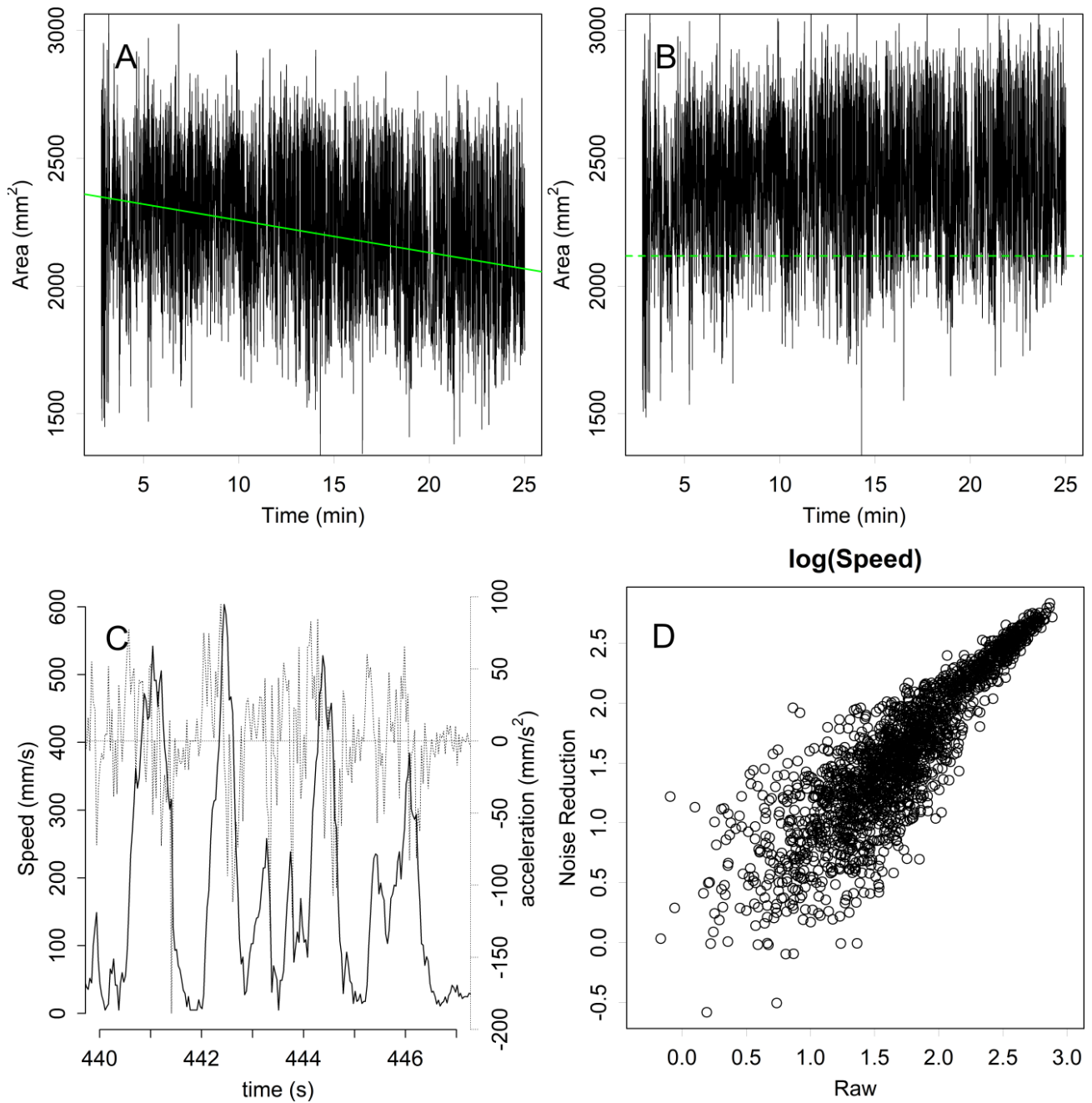

**Fig. S1**. **A**. Drift in the area with time. The linear relationship is indicated by the green line. The drift was caused by an artefact of exposure. **B**. After correction of the drift. The green line represents $\mu - 0.6\sigma$, which was used as an indicator of rearing. **C**. Speeds and accelerations in the raw data, without noise reduction (cf. Fig. 1C). The distributions were determined using the raw data. **D**. Comparison of the data with and without noise reduction. The effect was more significant for the slower speeds.

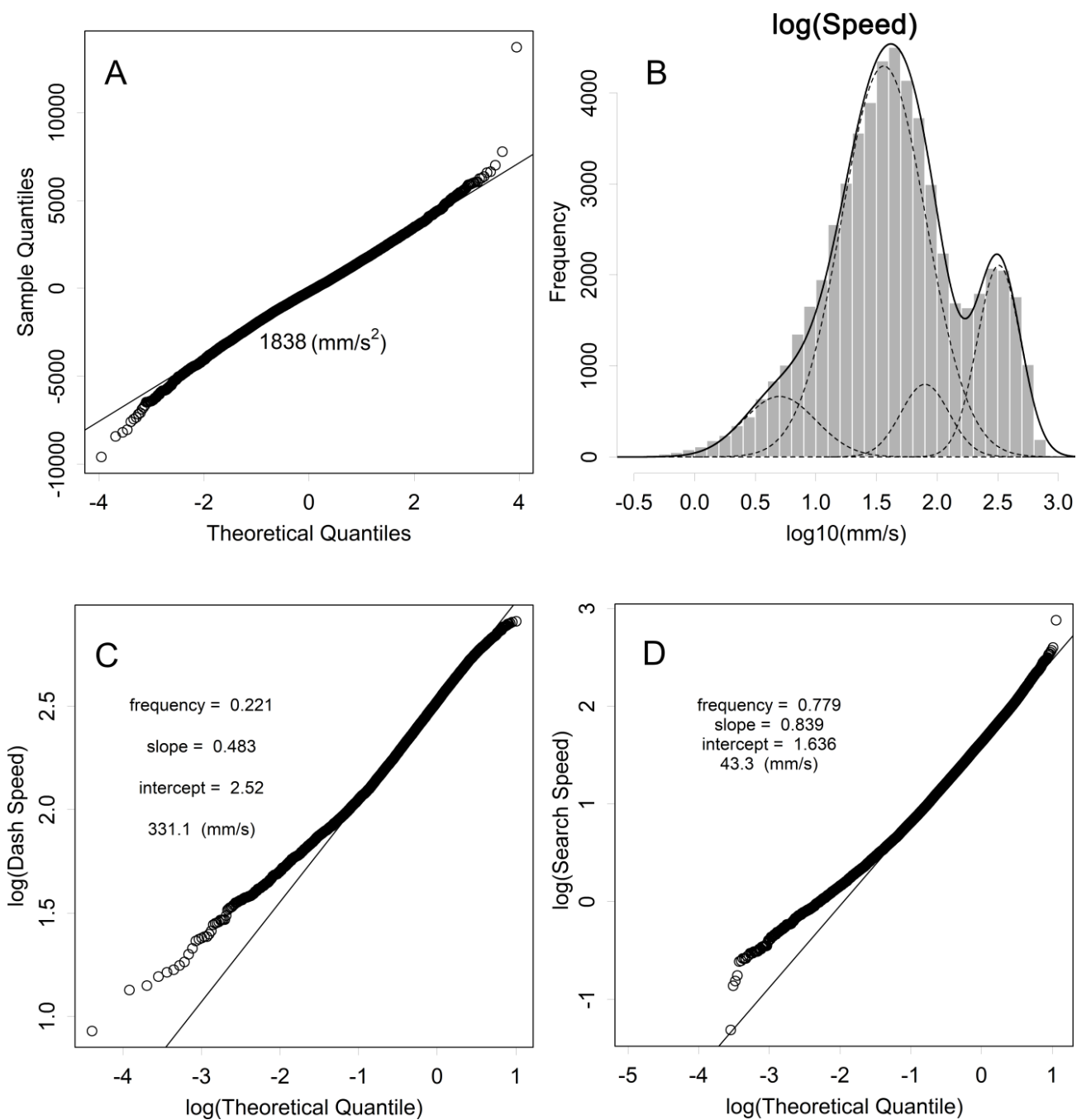

**Fig. S2**. **A**. Normal Q–Q plot of the dash accelerations. A linear relationship is obvious, with the slope indicating the $\sigma$ parameter. **B**. Fitting of the speeds using four lognormal distributions. **C**, **D**. Q–Q plot of the logarithms of the dash (**C**) and search (**D**) speeds against the logarithms of the exponential distribution. The estimation of the linear fitting was performed using the top 40% of the data, as the bottom data would contain data from the other cluster, which is an error that was caused

by the separation of the clusters using a line (Fig. 1D). Fig. 2B was drawn using the parameters obtained from these linear fittings.

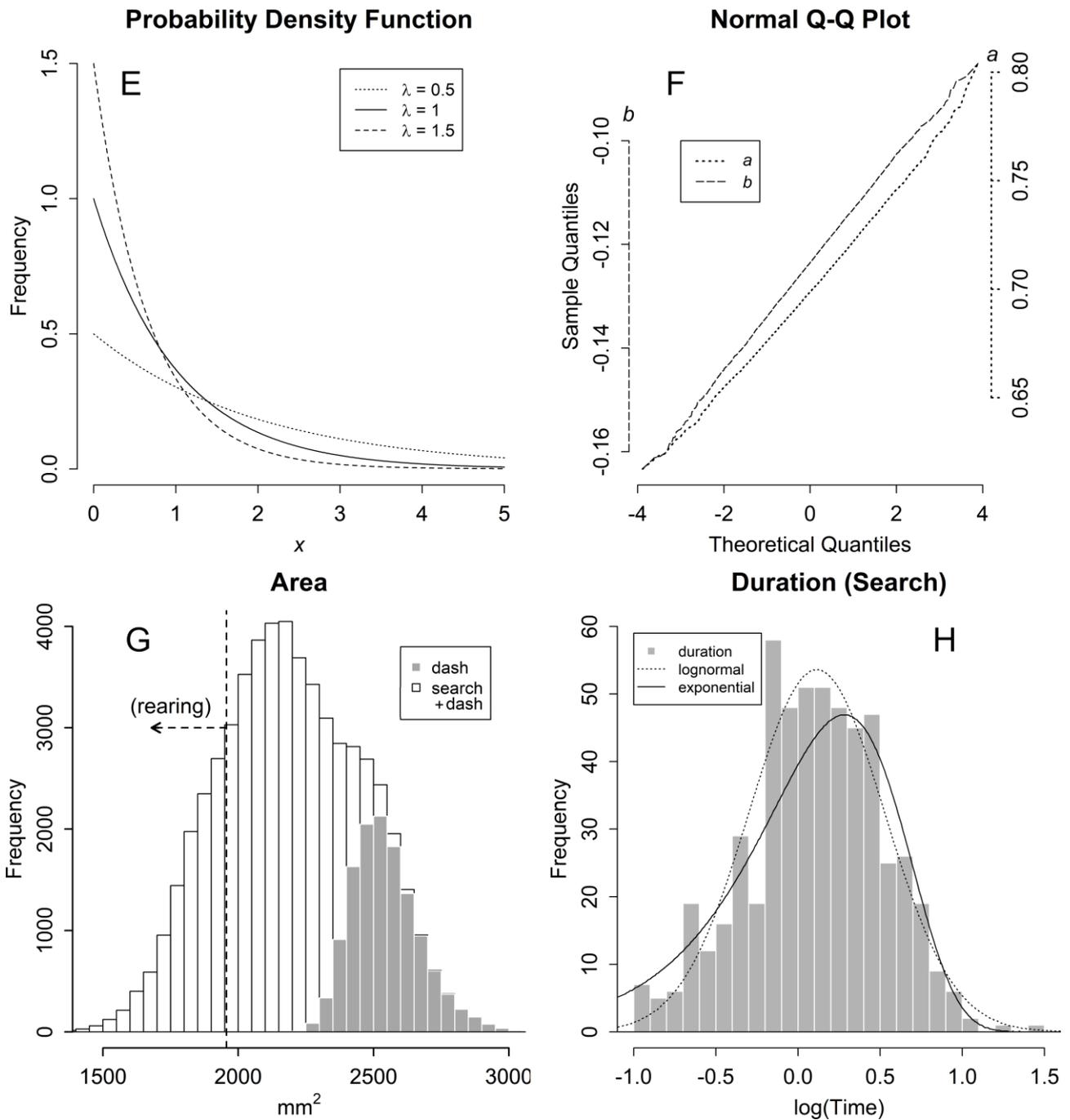

**Fig. S2**. **E**. Probability density function of the exponential distribution with different rates, $\lambda$. **F**. Distribution of the sampling errors that appeared in the estimations of *a* (F) and *b* (G). One hundred exponential random numbers were generated as $a = 0.5$ and $\lambda = 1.5$, and their logarithms were compared against the logarithms of the exponential distribution, to estimate *a* and *b* as the slope and intercept of the linear relationship, respectively. This was repeated 10,000 times, to obtain the distribution of sampling errors. **G**. Histogram of the projected areas of the animal (dash is coloured in grey). The vertical dotted line represents $\mu - 0.6\sigma$ of the search cluster, which was used as the indicator of rearing. **H**. Duration (log-scale) of the search clusters. Fittings using a lognormal distribution and an exponential distribution (log-scale) are shown (results obtained for the dash cluster are shown in Fig. 2D).

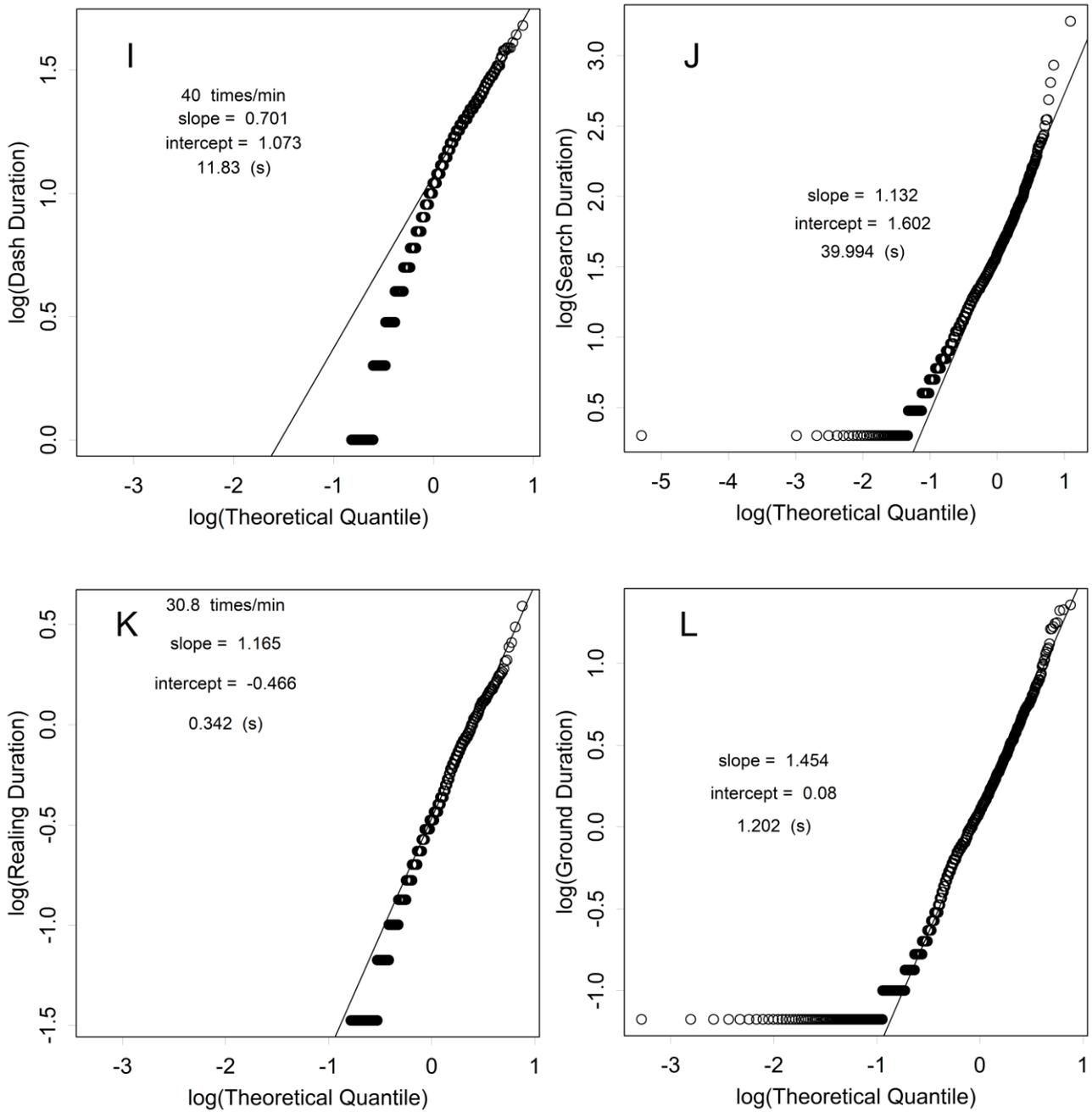

**Fig. S2. I–L**. Q–Q plot of the logarithms of the duration of dash (**I**), search (**J**), rearing (**K**), and grounding (**L**) against the logarithms of the exponential distribution. The separation of the clusters using a line (Fig. 1D) produced an error that generated very short fragments of sequences, such as is apparent at the bottom of the plots (J and L).

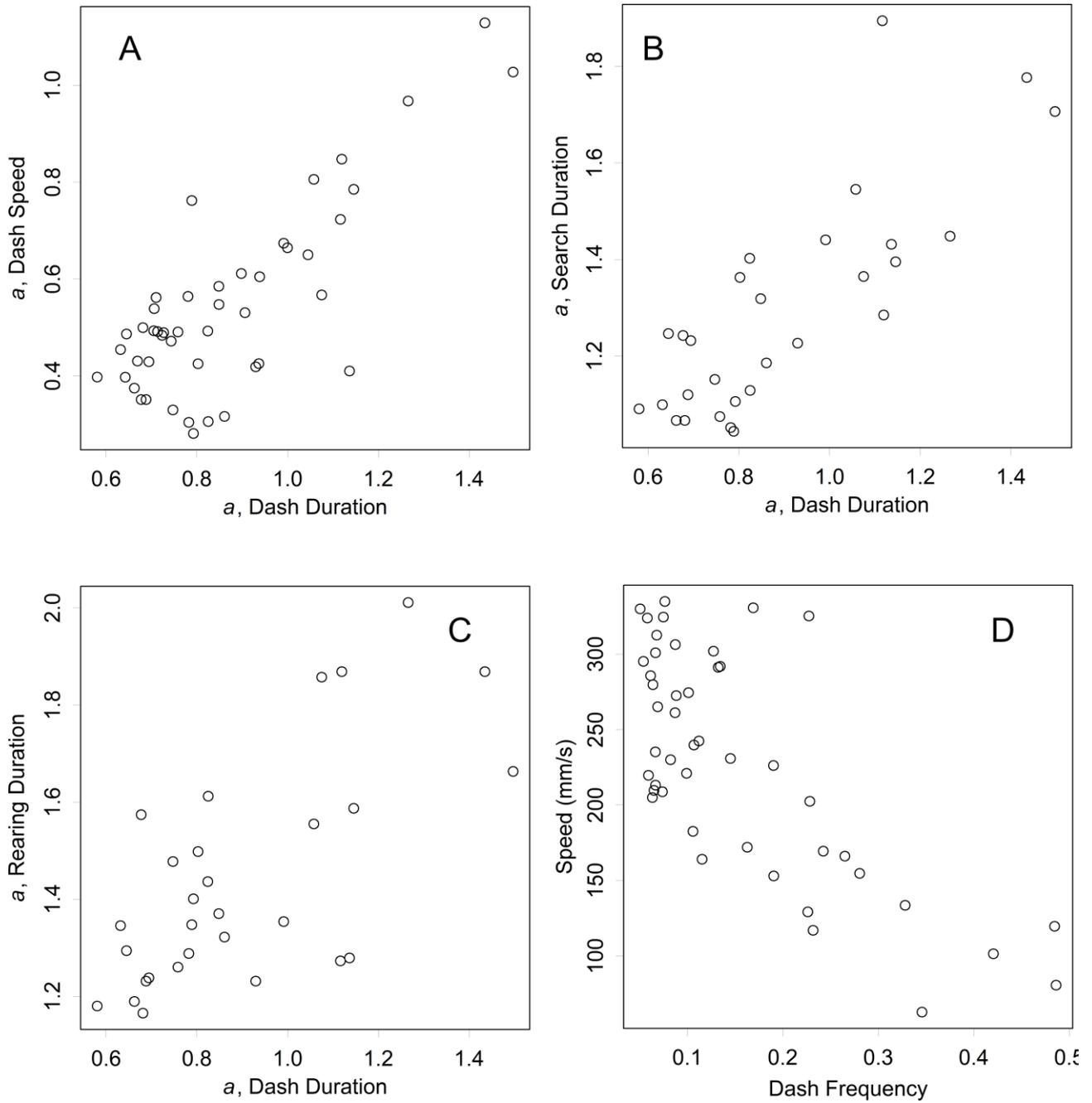

**Fig. S3. A–C**. Estimation of *a* in the forms of behaviours. **A**. Dash duration *vs* dash speed. **B**. Dash duration *vs* search duration. **C**. Dash duration *vs* rearing duration. **D**. Dash frequency *vs* dash speed. A weak negative correlation was observed.

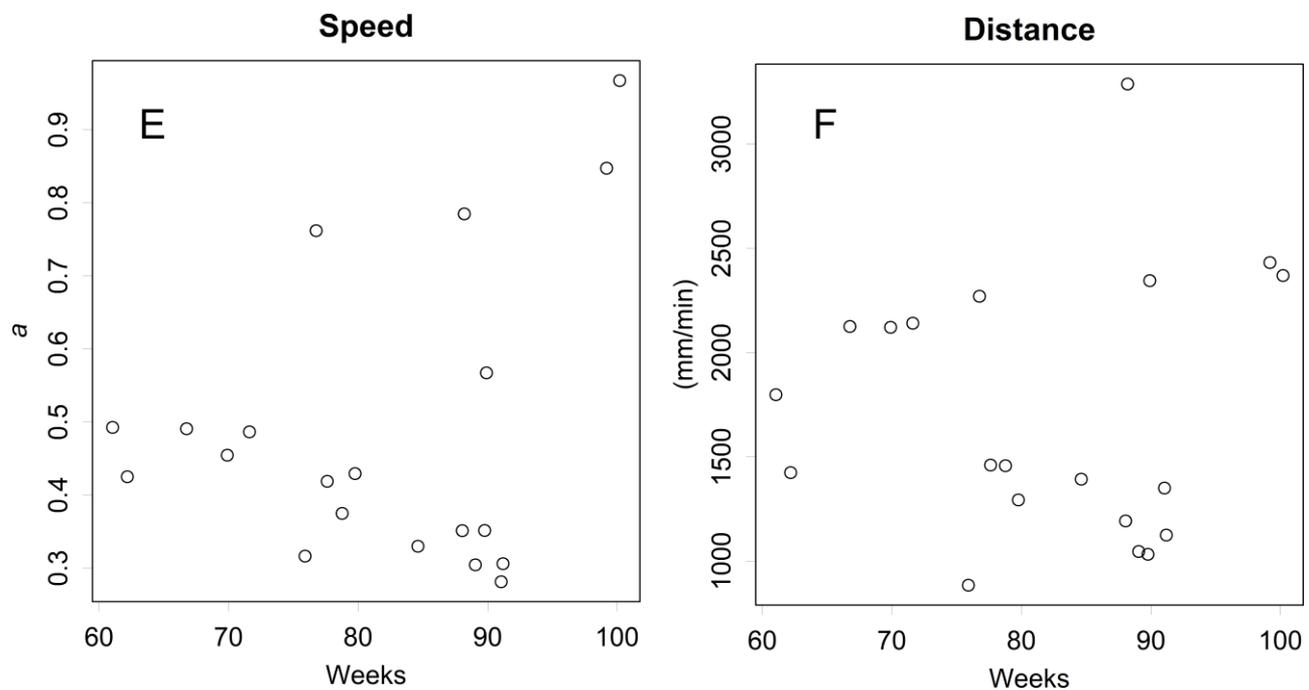

**Fig. S3. E**. Annual changes in *a*. (cf. those of speed in Fig. 3D). **F**. Annual changes in migrating distances over 1 min.

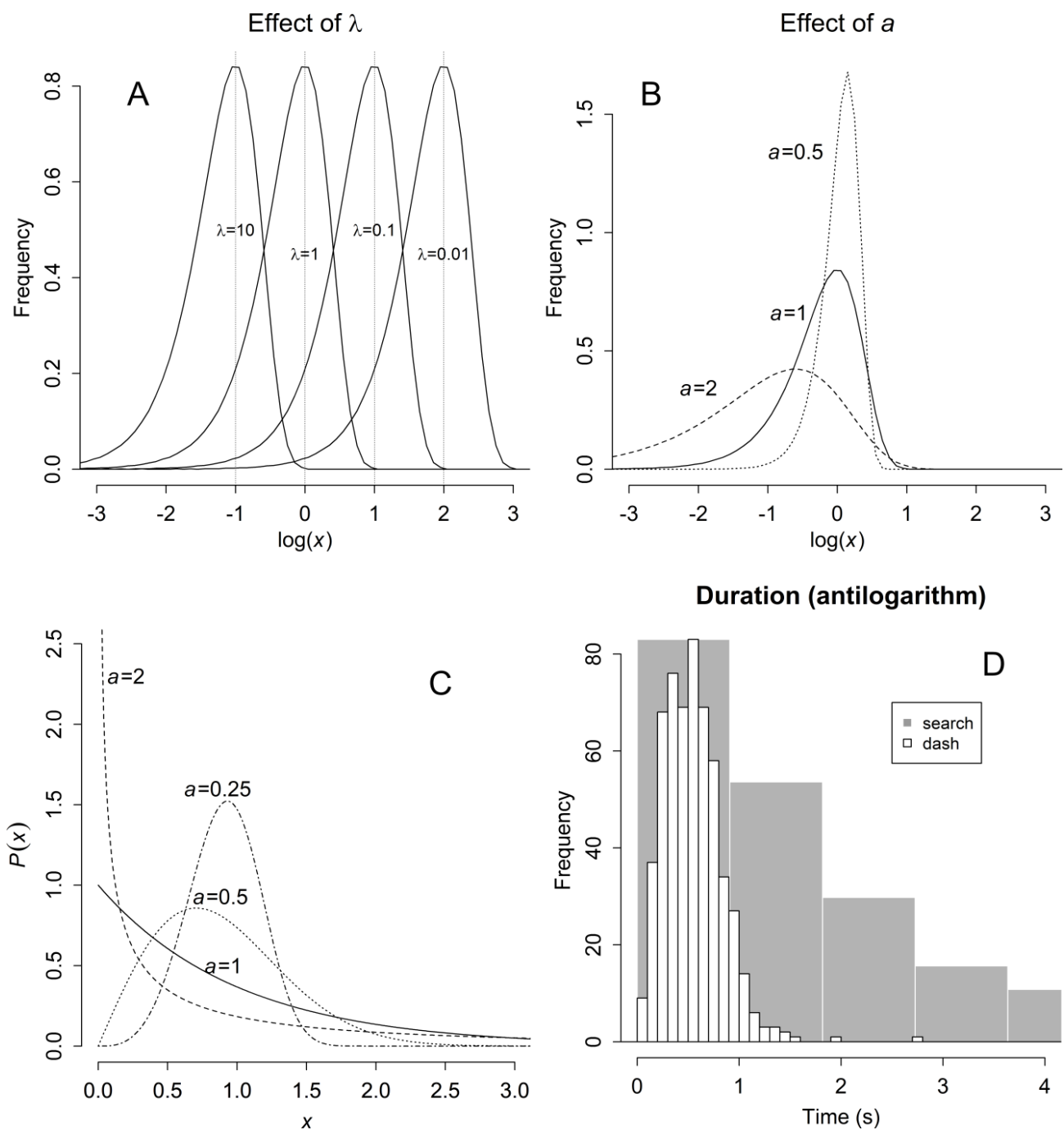

**Fig. S4. A**. Probability density function of the logarithms of the exponential distribution. Note that $\lambda$ changes the location of the mode at $\log(1/\lambda)$. **B**, **C**. Probability density function of the logarithms (**B**) and antilogarithms (**C**) of the exponential of the exponential distribution ($\lambda = 1$). The parameter *a* changes the shapes. **D.** Histograms of the duration of dash and search. The estimated *a* parameter for dash and search was 0.6 and 1.1, respectively.

**Experimental protocol used for movie processing**

0. A compact digital camera recorded a movie in .MOV format, which is not acceptable by the ImageJ analytical program. This format can be transferred to .AVI format using a conversion software and we used XMedia Recorde [15] for this purpose.

1. The movie was read by ImageJ using the AVI Reader function (dragging and dropping the movie file to the ImageJ window will start this function). Check the boxes "Use Virtual Stack" and "Convert to Grayscale". ImageJ reads the movie as a stack of discrete images.

2. Duplicate the movie. This treatment is required to alter the movie.

    Image > Duplicate "Duplicate stack"

3. To clarify the positions of the animal, each image should be simplified. To do this, subtract the background and convert the image into a binary white and black image. The background can be found as an average image of the movie.

    Image > Stacks > Z Project "Average Intensity"

4. The background is subtracted from each image of the duplicated movie. For a black mouse in a white box, inverted images gave a better result in the subtraction.

    (inversion) Edit > Invert (to both the duplicated movie and the average image)

    (subtraction) Process > Image Calculator> Operation > "Subtract"

    Image1: the inverted movie; Image 2: the inverted average image.

5. The subtracted movie has a black even background and the animal is shown in continuous grey. The image of the animal is binarized as:

    Image > Adjust > Threshold "Dark Background".

6. The position of the animal is then clarified by finding the centre of mass.

    Analyse > Set Measurements "Area", "Centre of mass", and "Display label".

    Analyse > Analyse Particles "Display results" "Clear results"

7. A spreadsheet is produced. This can be saved as a text file.

In the text file, extract the timing of each image from the "Label" column; this may be achieved by replacing " " and ":" with "\t". Generate another text file that has information pertaining to time, area, x, and y, to be read by the R program.